\documentclass[%
  reprint,
  showpacs,preprintnumbers,
  amsmath,amssymb,
  aps,
prl,
]{revtex4-1}

\usepackage{graphicx}
\usepackage{dcolumn}
\usepackage{bm}
\usepackage{hyperref}
\usepackage{multirow}
\usepackage{braket}
\usepackage{color}
\usepackage{ulem}

\usepackage{times}
\usepackage{booktabs}

\begin{document}

\preprint{APS/123-QED}

\title{
Phase Shift in Skyrmion Crystals 
}

\author{Satoru Hayami$^1$, Tsuyoshi Okubo$^2$, and Yukitoshi Motome$^1$}
\affiliation{
$^1$Department of Applied Physics, The University of Tokyo, Tokyo 113-8656, Japan \\
$^2$Department of Physics, The University of Tokyo, Tokyo 113-0033, Japan
}
 
\begin{abstract}
The skyrmion crystal is a periodic array of a swirling topological spin texture. 
Since it is regarded as an interference pattern by multiple helical spin density waves, the texture changes with the relative phases among the constituent waves. 
Although the phase degree of freedom is relevant to not only magnetism but also transport properties, its effect has not been elucidated thus far. 
We here theoretically show that a phase shift can occur in the skyrmion crystals to stabilize tetra-axial vortex crystals with staggered scalar spin chirality. 
This leads to spontaneous breaking of the lattice symmetry, which results in nonreciprocal transport phenomena even without the relativistic spin-orbit coupling. 
We show that such a phase shift is driven by long-range chirality interactions or thermal fluctuations in spin-charge coupled systems. 
\end{abstract}
\maketitle

{\it Introduction.---}
Skyrmion is a topological configuration of a continuous field
~\cite{Bogdanov89,Bogdanov94,rossler2006spontaneous}. 
Although it was originally proposed to explain hadrons in the particle theory~\cite{skyrme1962unified}, it has turned out to be realized in various forms in condensed matter physics~\cite{nagaosa2013topological}. 
One possible realization was discovered in magnets, in the form of a peculiar magnetic order called the skyrmion crystal (SkX)~\cite{Muhlbauer_2009skyrmion,yu2010real}.
The SkX is a periodic array of the skyrmion-like magnetic textures, which is regarded as an interference pattern by multiple helical spin density waves. 
It has attracted enormous attention since the swirling magnetic texture generates an emergent magnetic field through the Berry phase mechanism and gives rise to peculiar transport phenomena, such as the topological Hall effect~\cite{nagaosa2013topological,Lee_PhysRevLett.102.186601,Neubauer_PhysRevLett.102.186602}.

\begin{figure}[t!]
\begin{center}
\includegraphics[width=0.8 \hsize]{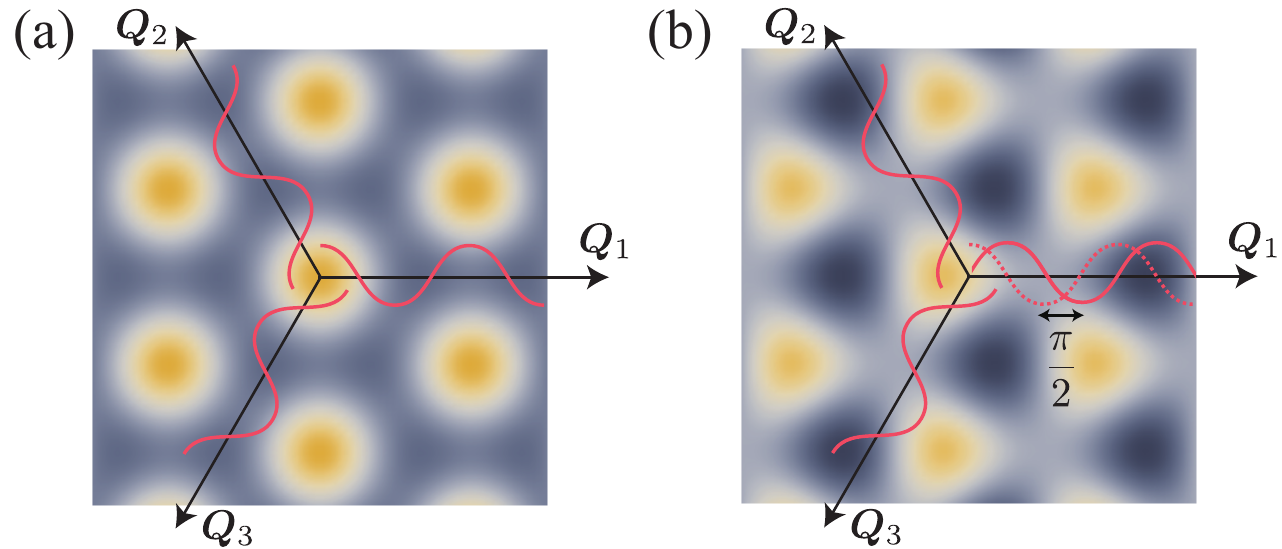} 
\caption{
\label{fig:moire}
(a) An interference pattern by the three scalar waves with the wave vectors $\bm{Q}_1$, $\bm{Q}_2$, and $\bm{Q}_3$. 
(b) The pattern by shifting the phase of $\bm{Q}_1$ by $\pi/2$, which breaks sixfold rotational symmetry.
}
\end{center}
\end{figure}

A single magnetic skyrmion is characterized by three quantities: skyrmion number, vorticity, and helicity~\cite{nagaosa2013topological}. 
This is also the case for the SkX: 
The spin texture in each magnetic unit cell can be specified by the three quantities. 
However, as the SkX is an interference pattern, it has another degree of freedom, the relative phases between the constituent waves. 
This is exemplified for three scalar waves in Fig.~\ref{fig:moire}: A shift of the relative phases leads to a different interference pattern. 
Such a phase degree of freedom exists in all the interference phenomena, except for those by linearly independent wave vectors. 
In addition, the SkX appears not in a continuous space but on a discrete lattice, which leads to a further variety of the interference patterns, even for the linearly independent waves.
The SkXs with different relative phases lead to different textures of the emergent magnetic field, and hence, different transport properties, but such an interesting possibility has not been elucidated thus far.

\begin{figure}[t!]
\begin{center}
\includegraphics[width=1.0 \hsize]{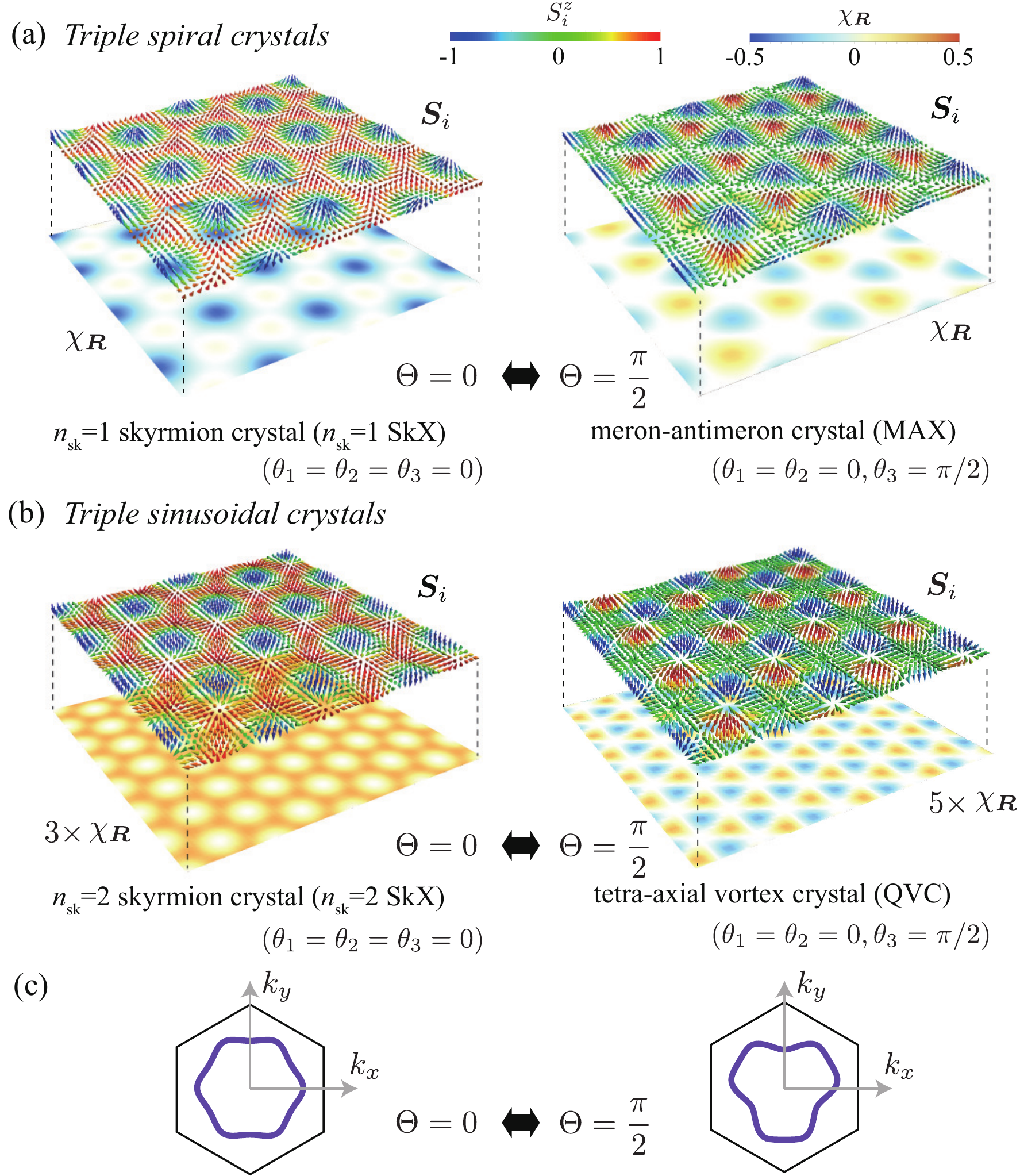} 
\caption{
\label{Fig:ponti}
Classification of the interference pattens. 
(a) Triple spiral crystals [Eq.~(\ref{eq:nsk1})]: 
(left) the $n_{\rm sk}=1$ skyrmion crystal ($n_{\rm sk}=1$ SkX) at $\Theta=0$ and 
(right) the meron-antimeron crystal (MAX) at $\Theta=\pi/2$. 
(b) Triple sinusoidal crystals [Eq.~(\ref{eq:nsk2})]: 
(left) the $n_{\rm sk}=2$ skyrmion crystal ($n_{\rm sk}=2$ SkX) at $\Theta=0$ and 
(right) the tetra-axial vortex crystal (TVC) at $\Theta=\pi/2$. 
In each figure, the upper and lower planes show the textures of spin $\bm{S}_i =(S_i^x, S_i^y, S_i^z)$ and the scalar spin chirality $\chi_{\bm{R}}$, respectively; the colors represent $S_i^z$ and $\chi_{\bm{R}}$. 
(c) The schematic Fermi surfaces at $\Theta=0$ and $\Theta=\pi/2$. 
}
\end{center}
\end{figure}

In this Letter, we theoretically propose a phase shift in the SkXs on the basis of phenomenological and numerical analyses. 
Starting from an effective spin model on a triangular lattice, we find that a chirality interaction in momentum space gives rise to a topological phase transition by the phase shift from a SkX to a tetra-axial vortex crystal (TVC) with a different texture of the scalar spin chirality. 
We show that the phase shift breaks sixfold rotational symmetry, leading to a spontaneous deformation of the Fermi surfaces and nonreciprocal transport even in the absence of the relativistic spin-orbit coupling. 
We also find that thermal fluctuations induce a similar phase shift in the Kondo lattice model through an entropically-driven chirality interaction. Our results open another route to control the magnetic textures through the phase shift in skyrmion-hosting materials. 

{\it Phase in skyrmion crystals.---}
Let us start by classifying noncoplanar spin textures according to the type of constituent waves and the relative phases. 
The first category is a superposition of spiral spin textures, which is exemplified by 
\begin{eqnarray}
\label{eq:nsk1}
\bm{S}_i =
\sum_{\nu=1}^3
\left(
       \sin \mathcal{Q}_\nu \cos \phi_\nu, 
       \sin \mathcal{Q}_\nu \sin \phi_\nu,
    - \cos \mathcal{Q}_\nu
  \right),
\end{eqnarray}
where $\mathcal{Q}_\nu=\bm{Q}_\nu \cdot \bm{r}_i+\theta_\nu$  and $\phi_\nu=\frac23 \pi(\nu-1)$. 
$\bm{Q}_\nu$ and $\theta_\nu$ are the ordering vector and phase for the $\nu$th spiral, respectively; 
$\bm{r}_i$ is the position vector for site $i$. 
In the following analyses, we choose $\bm{Q}_1=(Q,0)$, $\bm{Q}_2=(-Q/2,\sqrt{3}Q/2)$, and $\bm{Q}_3=(-Q/2,-\sqrt{3}Q/2)$ with the spiral pitch $Q$: 
$\bm{Q}_1+\bm{Q}_2+\bm{Q}_3=0$.  
The spin pattern in Eq.~(\ref{eq:nsk1}) changes with the total phase $\Theta=\sum_\nu \theta_\nu$ rather than each $\theta_\nu$. 
The change in $\Theta$ also modulates the texture of scalar spin chirality defined as $\chi_{\bm{R}}=\bm{S}_i \cdot (\bm{S}_j \times \bm{S}_k)$, where $\bm{R}$ represents the position vector at the center of a triangle with sites $i,j,k$ in the counterclockwise order. 
In Fig.~\ref{Fig:ponti}(a), we exemplify the spin textures at $\Theta=0$ and $\pi/2$ that correspond to the SkX with 
the skyrmion number of one  
$n_{\rm sk}=1$ ($n_{\rm sk}=1$ SkX) and the meron-antimeron crystal (MAX), respectively. 
The former retains sixfold rotational symmetry in $\chi_{\bm{R}}$, whose integration gives a nonzero net scalar chirality, but the latter lowers the symmetry to threefold with staggered $\chi_{\bm{R}}$, whose integration vanishes. 

The second category is a superposition of sinusoidal waves~\cite{Martin_PhysRevLett.101.156402,Hayami_PhysRevB.94.024424,Ozawa_PhysRevLett.118.147205}, which is exemplified by 
\begin{eqnarray}
\label{eq:nsk2}
\bm{S}_i =
\left(
    \cos \mathcal{Q}_1, 
     \cos \mathcal{Q}_2,  
    \cos \mathcal{Q}_3
  \right). 
\end{eqnarray}
Similar to Eq.~(\ref{eq:nsk1}), different $\Theta$ gives different spin and chirality textures, as shown in Fig.~\ref{Fig:ponti}(b) 
(the spin frame is rotated for better visibility). 
The spin texture with $\Theta=0$ is the SkX with $n_{\rm sk}=2$ ($n_{\rm sk}=2$ SkX), where $\chi_{\bm{R}}$ is sixfold symmetric with nonzero net scalar chirality, similar to the $n_{\rm sk}=1$ SkX in Fig.~\ref{Fig:ponti}(a). 
This type of SkX was found in the Kondo lattice model on a triangular lattice~\cite{Ozawa_PhysRevLett.118.147205}. 
The phase shift by $\pi/2$ in the $n_{\rm sk}=2$ SkX gives rise to a staggered threefold $\chi_{\bm{R}}$ 
breaking sixfold rotational symmetry without net scalar chirality, similar to the MAX in Fig.~\ref{Fig:ponti}(a). 
We find that the spin texture is given by a periodic array of four types of vortices; 
the vortex axes, which are defined by the volticity for $xy$, $yz$, and $zx$ components of spins as ($l_{xy}$, $l_{yz}$, $l_{zx}$), point to four corners of the tetrahedron~\cite{SM_QVC}. 
Hence, we call the $\Theta=\pi/2$ state TVC. 

Due to the cancellation of $\chi_{\bm{R}}$, MAX and TVC do not lead to the topological Hall effect, in contrast to the SkXs. 
Nevertheless, the breaking of sixfold rotational symmetry leads to deformations of the Fermi surfaces, as schematically shown in Fig.~\ref{Fig:ponti}(c)~\cite{comment_FS_nsk1SkX}. 
Such band deformations can induce direction-dependent nonreciprocal transport phenomena, as in spin-orbit-coupled metals with toroidal moments~\cite{gorbatsevich1994toroidal,Hayami_PhysRevB.90.024432,Gao_PhysRevLett.122.227402}. 
However, the qualitative difference lies in the fact that the present mechanism works even without the spin-orbit coupling~\cite{ishizuka2019anomalous, hayami2020spontaneous}. 

The optimal $\Theta$ will be determined by multiple factors, such as lattice geometry and exchange interactions between the spins. 
In the previous studies, the SkXs with $\Theta=0$ are stabilized, e.g, by the Dzyaloshinskii-Moriya (DM)~\cite{rossler2006spontaneous,Yi_PhysRevB.80.054416}, four-spin~\cite{heinze2011spontaneous,brinker2019chiral,Laszloffy_PhysRevB.99.184430,paul2019role}, frustrated~\cite{Okubo_PhysRevLett.108.017206,leonov2015multiply,Lin_PhysRevB.93.064430}, and spin-charge interactions~\cite{Ozawa_PhysRevLett.118.147205,Hayami_PhysRevB.99.094420,wang2019skyrmion} on a variety of lattices under the constraint on the fixed spin length at each site. 
The key question addressed here is whether there is any exchange interaction that directly controls $\Theta$. 
In the following, we show that a particular type of multiple-spin interaction, which naturally arises in itinerant magnets, is relevant to the phase shift. 

{\it Effective spin model.---}
We first demonstrate a phase shift in an effective spin model with long-range interactions derived from the Kondo lattice model on a triangular lattice [introduced later in Eq.~(\ref{eq:Ham})]. 
The Hamiltonian is given by 
\begin{eqnarray}
\label{eq:Spin_Model}
\mathcal{H}= & 2\sum_{\nu=1}^3
\left[ -J\bm{S}_{\bm{Q}_{\nu}} \cdot \bm{S}_{-\bm{Q}_{\nu}}
+\tilde{K} (\bm{S}_{\bm{Q}_{\nu}} \cdot \bm{S}_{-\bm{Q}_{\nu}})^2 \right] \nonumber \\
&+ \tilde{L} \left[ \left\{\bm{S}_{\bm{Q}_1}\cdot (\bm{S}_{\bm{Q}_2} \times \bm{S}_{\bm{Q}_3}) \right\}^2+
{\rm H.c. }
\right], 
\end{eqnarray}
where the magnetic interactions are defined for the same set $\bm{Q}_{\nu}$ with $Q=\pi/3$ that are dictated by the sixfold-symmetric Fermi surface nesting (we take the lattice constant unity). 
The first two terms describe bilinear and biquadratic interactions for the Fourier component of spins, $\bm{S}_{\bm{Q}_\nu} = (1/\sqrt{N}) \sum_i \bm{S}_i e^{i\bm{Q}_\nu\cdot\bm{r}_r}$, which are derived by second- and fourth-order perturbation expansions in terms of the spin-charge coupling, respectively~\cite{Hayami_PhysRevB.95.224424}; $J>0$ and $\tilde{K}=K/N>0$, and $N$ denotes the number of sites. 
Meanwhile, the third term with $\tilde{L}=L/N^2$ represents an interaction between the scalar spin chirality composed of $\bm{S}_{\bm{Q}_\nu}$. 
This appears in the sixth-order perturbation, among many other terms. 
It is worthy to note that this long-range chirality interaction is the lowest-order contribution whose energy depends on $\Theta$ under $\sum_\nu \bm{Q}_\nu=0$. 
This is in stark contrast to the short-range 
chirality interaction that drives a transition from a coplanar single-$Q$ to a noncoplanar multiple-$Q$ state, rather than the phase shift~\cite{grytsiuk2020topological}. 

\begin{figure}[t!]
\begin{center}
\includegraphics[width=1.0 \hsize]{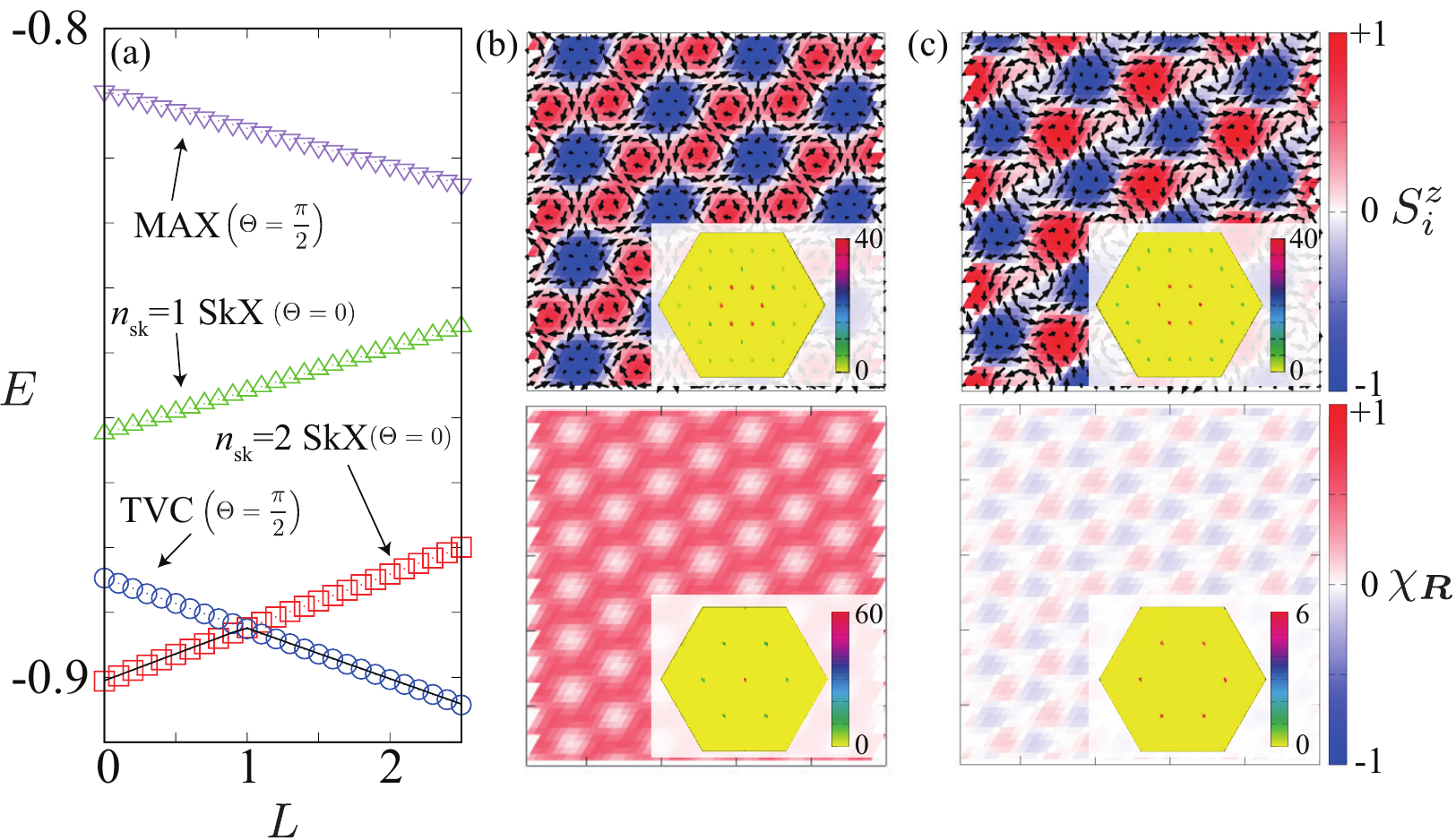} 
\caption{
\label{Fig:Spin_model}
(a) $L$ dependence of the variational energy per site, $E$, at $J=1$ and $K=0.4$ for the model in Eq.~(\ref{eq:Spin_Model}). 
The solid line shows the result by the simulated annealing at temperature $T=10^{-4}$ for $N=96^2$. 
(b), (c) (Upper panels) Spin configurations in (b) the $n_{\rm sk}=2$ SkX at $L=0$ and (c) the TVC at $L=2$ obtained by the simulated annealing. 
The color scale indicates $S_i^z$, while the arrows indicate $(S_i^x, S_i^y)$. 
(Lower panels) Corresponding chirality textures. 
Each inset displays the square root of the structure factor, $\sqrt{S_s(\bm{q})}$ or $\sqrt{S_\chi(\bm{q})}$. 
}
\end{center}
\end{figure}

To clarify the effect of the chirality interaction, we investigate the ground-state phase diagram of the model in Eq.~(\ref{eq:Spin_Model}) by variational calculations. 
We compare the energy of the four states shown in Fig.~\ref{Fig:ponti} while varying three $\theta_\nu$ as the variational parameters under the constraint $|\bm{S}_i|=1$ at each site. 
Figure~\ref{Fig:Spin_model}(a) shows the result for $J=1$ and $K=0.4$ while changing $L$ for the system with $N=96^2$ sites under the periodic boundary conditions. 
We find that the energies for the triple sinusoidal crystals in Eq.~(\ref{eq:nsk2}) are lower than those for the triple spiral crystals in Eq.~(\ref{eq:nsk1})~\cite{comment_SkX_ziba}. 
Between the triple sinusoidal crystals, the $n_{\rm sk}=2$ SkX has a lower energy than the TVC for $0\leq L\lesssim 1$~\cite{Hayami_PhysRevB.95.224424}, while the latter wins for larger $L$. 
We find that the optimal $\theta_\nu$ are obtained as $(\theta_1, \theta_2, \theta_3) = (\pi/3, -\pi/3, 0)$ for the former and $(\pi/3, \pi/6, 0)$ for the latter (any permutation is allowed). 
These values are selected by spin length normalization and discreteness of the lattice so that the skyrmion or vortex cores locate at the triangle centers. 
In addition, we also find that, for the energetically-higher triple spiral crystals, the $n_{\rm sk}=1$ SkX is taken over by the MAX in the larger $L$ region.
Thus, our results clearly show that the long-range chirality interaction prefers the $\Theta=\pi/2$ states to the $\Theta=0$ ones, namely, it brings about the phase shift in SkXs. 

The variational results are confirmed by the simulated annealing following the manner in Ref.~\cite{Hayami_PhysRevB.95.224424}; the obtained energy for $N=96^2$ is plotted as the solid curve in Fig.~\ref{Fig:Spin_model}(a)~\cite{SM_QVC}. 
The typical spin and chirality textures obtained by the simulated annealing are shown in Figs.~\ref{Fig:Spin_model}(b) and \ref{Fig:Spin_model}(c), 
which are basically the same as those in Figs.~\ref{Fig:ponti}(a) and \ref{Fig:ponti}(b), respectively. 
We also plot the spin and chirality structure factors, $S_s(\bm{q})=(1/N) \sum_{\alpha=x,y,z}\sum_{j,l} S_j^{\alpha} S_l^{\alpha} e^{i \bm{q}\cdot (\bm{r}_j-\bm{r}_l)}$ and $S_{\chi}(\bm{q})=(1/N)\sum_{\mu}\sum_{\bm{R},\bm{R}' \in \mu} \chi_{\bm{R}}\chi_{\bm{R}'} e^{i \bm{q}\cdot (\bm{R}-\bm{R}')}$, respectively, in each inset. 
For the latter,  
$\mu=(u, d)$ represent upward and downward triangles, respectively. 
The two states share the peaks at $S_s(\bm{Q}_\nu)$ and $S_\chi(2\bm{Q}_\nu)$, while only the former exhibits the peaks at $S_s(\bm{Q}_1+\bm{Q}_2)$ and $S_\chi(0)$.

{\it Kondo lattice model.---}
The analysis of the effective model illuminates a relevant interaction to control $\Theta$. 
Below we demonstrate that a similar phase shift is also caused by thermal fluctuations, by considering the Kondo lattice model, whose Hamiltonian is given by  
\begin{eqnarray}
\label{eq:Ham}
\mathcal{H} = 
-\sum_{i, j,  \sigma} t_{ij} c^{\dagger}_{i\sigma}c_{j \sigma}
+J_{\rm K} \sum_{i} \bm{s}_i \cdot \bm{S}_i - H \sum_i S_i^z. 
\end{eqnarray}
The first term represents the kinetic energy of itinerant electrons, where $c^{\dagger}_{i\sigma}$ ($c_{i \sigma}$) is a creation (annihilation) operator of an itinerant electron at site $i$ and spin $\sigma$. 
The second term represents the exchange coupling between itinerant electron spins $\bm{s}_i=(1/2)\sum_{\sigma, \sigma'}c^{\dagger}_{i\sigma} \bm{\sigma}_{\sigma \sigma'} c_{i \sigma'}$ [$\bm{\sigma}=(\sigma^x,\sigma^y,\sigma^z)$ is the vector of Pauli matrices] and classical localized spins $\bm{S}_i$ with $|\bm{S}_i|=1$. 
The third term represents the Zeeman coupling to an external magnetic field $H$. 
The model parameters are common to those in the previous study for the ground state~\cite{Ozawa_PhysRevLett.118.147205}. 
We study the finite-temperature behavior of the model in Eq.~(\ref{eq:Ham}) by performing the 
Langevin dynamics simulations with the kernel polynomial method~\cite{Weis_RevModPhys.78.275,Barros_PhysRevB.88.235101,Barros_PhysRevB.90.245119,Ozawa_doi:10.7566/JPSJ.85.103703,Ozawa_PhysRevLett.118.147205,Wang_PhysRevLett.117.206601,Chern_PhysRevB.97.035120}.
The simulations are done for $N=96^2$ and $120^2$ sites, and the thermal averages are taken for 100-800 samplings.

\begin{figure}[t!]
\begin{center}
\includegraphics[width=1.0 \hsize]{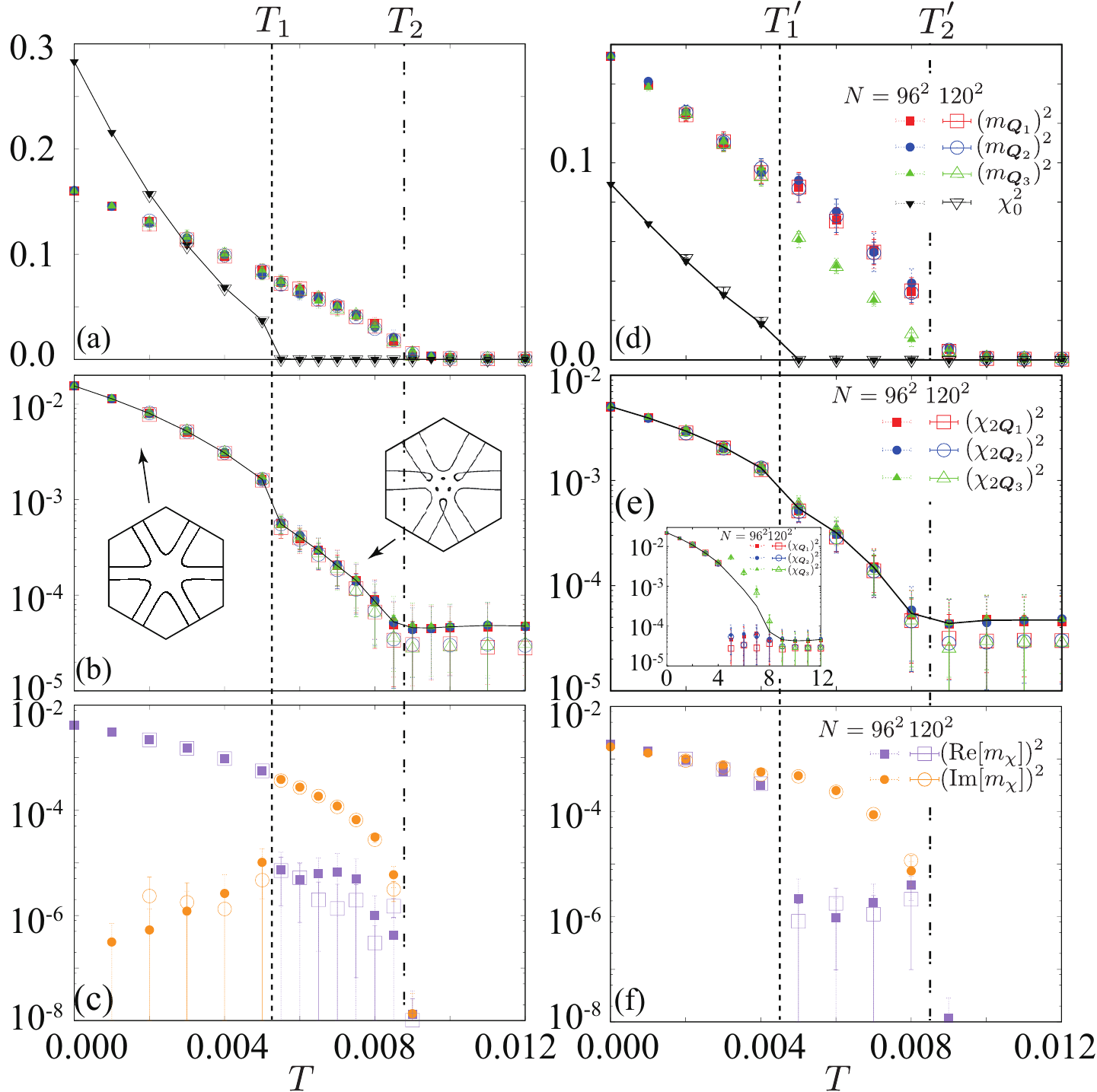} 
\caption{
\label{Fig:KLM_zeromag}
Temperature dependences of (a), (d) $m_{\bm{Q}_\nu}^{2}$ and $\chi_0^{2}$, (b), (e) $\chi_{2\bm{Q}_\nu}^{2}$, and (c), (f) Re$[m_\chi]^{2}$ and Im$[m_\chi]^{2}$. 
The data are calculated at $H=0$ in (a)-(c) and $H=0.004$ in (d)-(f). 
The dashed (dash-dotted) line represents the transition temperature $T_1$ ($T_2$) in (a)-(c) and $T'_1$ ($T'_2$) in (d)-(f).
The inset of (b) displays the Fermi surfaces in the $n_{\rm sk}=2$ SkX and TVC. 
The inset of (e) plots $\chi_{\bm{Q}_\nu}$. 
}
\end{center}
\end{figure}

Figures~\ref{Fig:KLM_zeromag}(a)-(c) show the finite-temperature results at $H=0$ where the ground state is the $n_{\rm sk}=2$ SkX~\cite{Ozawa_PhysRevLett.118.147205}. 
The $\bm{q}$ components of the magnetic moment, $m_{\bm{q}}=\sqrt{S_s(\bm{q})/N}$, at $\bm{q}=\bm{Q}_1$, $\bm{Q}_2$, and $\bm{Q}_3$, and the scalar chirality, $\chi_{\bm{q}}=\sqrt{S_{\chi}(\bm{q})/N}$, at $\bm{q}=0$ are plotted in Fig.~\ref{Fig:KLM_zeromag}(a). 
The data indicate two phase transitions at $T_1 \sim 0.0055$ and $T_2 \sim 0.009$. 
The transition at $T_1$ is characterized by the onset of $\chi_{0}$, suggesting that the $n_{\rm sk}=2$ SkX remains stable up to $T_1$, although the true magnetic long-range order is limited to zero temperature due to the Mermin-Wagner theorem~\cite{Mermin_PhysRevLett.17.1133}; the state for $0<T<T_1$ is a chiral spin liquid with a quasi-long-range spin texture of the $n_{\rm sk}=2$ SkX.
Meanwhile, the transition at $T_2$ appears to be signaled by $m_{\bm{Q}_\nu}$, but this is superficial due to the Mermin-Wagner theorem~\cite{Mermin_PhysRevLett.17.1133}. 
Instead, the transition is identified by the onset of the higher harmonics $\chi_{2\bm{Q}_\nu}$ as plotted in Fig.~\ref{Fig:KLM_zeromag}(b). 
At the same time, we find that the triple scalar product of $\bm{m}_{\bm{Q}_\nu}$, $m_{\chi}= \bm{m}_{\bm{Q}_1}\cdot (\bm{m}_{\bm{Q}_2}\times \bm{m}_{\bm{Q}_3})$ becomes nonzero as plotted in Fig.~\ref{Fig:KLM_zeromag}(c). 
Interestingly, $m_\chi$ is a pure imaginary number for $T_1<T<T_2$, while it becomes purely real below $T_1$. 
This indicates that a phase shift by $\pi/2$ occurs at $T_2$. 
From these results, we conclude that the low- and intermediate-temperature phases are the $n_{\rm sk}=2$ SkX and TVC, respectively, with quasi-long-range magnetic orders. 
Thus, the transition at $T_1$ is driven by the phase shift between the two chiral spin liquid states. 

The appearance of the TVC at finite temperature is explained by an effective chirality interaction as follows. 
At the mean-field level, the entropic contributions are in general given in the form of $n$th-order magnetic interactions as $T\sum_{\bm{q}_1 \cdots \bm{q}_n} (\bm{S}_{\bm{q}_1}\cdot \bm{S}_{\bm{q}_2})\cdots (\bm{S}_{\bm{q}_{n-1}}\cdot \bm{S}_{\bm{q}_n})\delta(\bm{q}_1+\cdots +\bm{q}_n)$~\cite{Reimers_PhysRevB.43.865,Okubo_PhysRevB.84.144432}. 
Among them, the lowest-order contribution to the phase shift at zero field appears in the sixth order. 
By considering Eq.~(\ref{eq:nsk2}), the entropic contribution is roughly estimated as 
\begin{eqnarray}
&T {\rm Re}[(\bm{S}_{\bm{Q}_1}\cdot \bm{S}_{\bm{Q}_1})(\bm{S}_{\bm{Q}_2}\cdot \bm{S}_{\bm{Q}_2})(\bm{S}_{\bm{Q}_3}\cdot \bm{S}_{\bm{Q}_3})] \nonumber \\
&= T {\rm Re}[\left\{\bm{S}_{\bm{Q}_1}\cdot (\bm{S}_{\bm{Q}_2} \times \bm{S}_{\bm{Q}_3}) \right\}^2] \propto T \cos 2 \Theta. 
\end{eqnarray}
Thus, the entropic term favors the TVC at finite temperature.

We display the Fermi surfaces in the TVC in the inset of Fig.~\ref{Fig:KLM_zeromag}(b), which is deformed by breaking sixfold rotational symmetry from the one in the SkX. 
The functional form of the deformation is described by $k_y (k_y^2-3k_x^2)$. 
This leads to nonreciprocal transports, even when the system does not include the spin-orbit coupling, as stated above.

Finally, let us discuss a similar phase shift for the $n_{\rm sk}=1$ SkX 
at nonzero field. 
Figures~\ref{Fig:KLM_zeromag}(d)-(f) summarize the results at $H=0.004$. 
Similar to the $H=0$ case in Figs.~\ref{Fig:KLM_zeromag}(a)-(c), there are two phase transitions at $T'_1 \sim 0.0045$ and $T'_2 \sim 0.0085$. 
The low-temperature state below $T'_1$ is the $n_{\rm sk}=1$ SkX (with quasi-long-range order in the $xy$ components)~\cite{Ozawa_PhysRevLett.118.147205}. 
Although the phase shift of $\pi/2$ from the $n_{\rm sk}=1$ SkX is expected to give rise to the MAX as shown in Fig.~\ref{Fig:ponti}(a), we find that the intermediate phase for $T'_1<T<T'_2$ is the TVC also in this case, as indicated by $\chi_0=0$, $\chi_{2\bm{Q}_\nu}\neq 0$, ${\rm Re}[m_\chi]=0$, and ${\rm Im}[m_\chi]\neq 0$. 
This is because the Zeeman energy gain in the MAX is not sufficient to overcome the TVC. 
We note, however, that the symmetry is further lowered by the magnetic field from the TVC at $H=0$; 
$m_{\bm{Q}_{3}}$ has different intensities from $m_{\bm{Q}_1}$ and $m_{\bm{Q}_2}$, reflecting the breaking of threefold rotational symmetry. 
This is also found in $\chi_{\bm{Q}_\nu}$ in the inset of Fig.~\ref{Fig:KLM_zeromag}(e). 
We note that similar symmetry breaking was also found in the high-field region at zero temperature~\cite{Ozawa_PhysRevLett.118.147205}. 

{\it Conclusion.---}
To summarize, we have theoretically guided a new direction of controlling the SkXs by using the phase degree of freedom among the constituent spin density waves. 
The phase shift turns the SkXs into the TVCs characterized by staggered emergent magnetoelectric fields and breaking of the lattice symmetry. 
We found that such a phase shift is driven by long-range chirality interactions or thermal fluctuations.

Our results indicate that the skyrmion-based physics can be switched on and off by changing the relative phases among the constituent waves. 
Moreover, the staggered align of the scalar spin chirality in the TVC induces the nonreciprocal transport even on a centrosymmetric lattice, which opens fertile physics related to ``chiraltronics". 
The phase shift in SkXs would be realized in the centrosymmetric skyrmion-hosting materials where the multiple-spin interactions rooted in the spin-charge coupling might play an important role~\cite{kurumaji2019skyrmion,hirschberger2019skyrmion,hirschberger2019topological,Ishiwata_PhysRevB.101.134406}. 
While it is not straightforward to distinguish the spin textures with different relative phases by the neutron diffraction and resonant x-ray scattering, our findings suggest that the angle-resolved photoemission spectroscopy and macroscopic nonreciprocal transport measurements will give good probes to detect the TVC. 

\begin{acknowledgments}
The authors thank R. Ozawa for the fruitful discussions. 
This research was supported by JSPS KAKENHI Grants Numbers JP18H04296 (J-Physics), JP18K13488, JP19K03752, JP19H01834, 19K03740, and JP19H05825, JST CREST (JP-MJCR18T2), and JST PRESTO (JP-MJPR1912). 
This work was also supported by the Toyota Riken Scholarship. 
Parts of the numerical calculations were performed in the supercomputing systems in ISSP, the University of Tokyo.
\end{acknowledgments}

\bibliographystyle{apsrev}
\bibliography{ref}

\begin{thebibliography}{45}
\bibitem[{com({\natexlab{a}})}]{comment_FS_nsk1SkX}
\bibinfo{note}{Although the inversion symmetry is broken also in the $n_{\rm
  sk}=1$ SkX, but the Fermi surfaces are still symmetric due to sixfold
  rotational symmetry. In this case, the asymmetric spin-momentum locking
  occurs.}
\bibitem[{com({\natexlab{b}})}]{comment_SkX_ziba}
\bibinfo{note}{In the presence of the magnetic field, the $n_{\rm sk}=1$ SkX
  can become the ground state. See also Fig.~4.}
\end{thebibliography}

\end{document}